\documentclass[twocolumn,showpacs,preprintnumbers,amsmath,amssymb]{revtex4}

\usepackage{epsfig}

\def\cal#1{{\cal #1}}

\def\m@th{\mathsurround=0pt}
\def\n@space{\nulldelimiterspace=0pt \m@th}%1
\def\biggg#1{{\mbox{$\left#1\vbox to 20.5pt{}\right.\n@space$}}}%2

\def\beginenum{\begin{enumerate}}
\def\endenum{\end{enumerate}}
\def\bitem{\begin{itemize}}
\def\eitem{\end{itemize}}
\def\bray{\begin{array}}
\def\eray{\end{array}}
\def\begindoc{\begin{document}}
\def\enddoc{\end{document}}
\def\bq{\begin{equation}}
\def\eq{\end{equation}}
\def\bqy{\begin{eqnarray}}
\def\eqy{\end{eqnarray}}
\def\bqyn{\begin{eqnarray*}}
\def\eqyn{\end{eqnarray*}}
\def\bc{\begin{center}}
\def\ec{\end{center}}
\def\bfll{\begin{flushleft}}
\def\efll{\end{flushleft}}
\def\bflr{\begin{flushright}}
\def\eflr{\end{flushright}}

\newcommand{\Avec}{\mbox{\boldmath $A$}}
\newcommand{\Bvec}{\mbox{\boldmath $B$}}
\newcommand{\Evec}{\mbox{\boldmath $E$}}
\newcommand{\Fvec}{\mbox{\boldmath $F$}}
\newcommand{\Gvec}{\mbox{\boldmath $G$}}
\newcommand{\Rvec}{\mbox{\boldmath $R$}}
\newcommand{\Uvec}{\mbox{\boldmath $U$}}
\newcommand{\Vvec}{\mbox{\boldmath $V$}}
\newcommand{\evec}{\mbox{\boldmath $e$}}
\newcommand{\jvec}{\mbox{\boldmath $j$}}
\newcommand{\kvec}{\mbox{\boldmath $k$}}
\newcommand{\nvec}{\mbox{\boldmath $n$}}
\newcommand{\uvec}{\mbox{\boldmath $u$}}
\newcommand{\vvec}{\mbox{\boldmath $v$}}
\newcommand{\wvec}{\mbox{\boldmath $w$}}
\newcommand{\xvec}{\mbox{\boldmath $x$}}
\newcommand{\omegavec}{\mbox{\boldmath $\omega$}}
\newcommand{\Omegavec}{\mbox{\boldmath $\Omega$}}

\begin{document}

\title{ Wave Localization and Density Bunching in Pair Ion Plasmas}
\author{Swadesh M. Mahajan}
\email{mahajan@mail.utexas.edu} \affiliation {Institute for Fusion
Studies, The University of Texas at Austin, Austin,Tx 78712}
\author{Nana L. Shatashvili} \email{shatash@ictp.it} \affiliation
{Faculty of Exact and Natural Sciences, Javakhishvili Tbilisi
State University, Tbilisi 0128, Georgia\\
Andronikashvili Institute of Physics, Tbilisi 0177, Georgia}

\begin{abstract}
{By investigating the nonlinear propagation of high intensity
electromagnetic (EM) waves in a pair ion plasma, whose symmetry is
broken via contamination by a small fraction of high mass immobile
ions, it is shown that this new and interesting state of
(laboratory created) matter is capable of supporting structures
that strongly localize and bunch the EM radiation with density
excess in the region of localization. Testing of this prediction
in controlled laboratory experiments can lend credence, $\it
inter$ $\it alia$, to conjectures on structure formation (via the
same mechanism) in the MEV era of the early universe.}

\end{abstract}

 \maketitle

\clearpage

The physics of pair plasmas was turned into an even more exciting
field of investigation when it descended from its astrophysical
heights to the terrestrial laboratory. The story of laboratory
pair plasmas had its defining moment with the successful creation
of a "sufficiently" dense
%, and relatively pure
pair-ion (pi) plasma -- consisting of equal-mass, positive and negative fullerene ions ($C_{60}^+$
and $C_{60}^-$) \cite{bib:OH1}. Unlike the electron--positron (e--p) plasma systems (both of  the
astrophysical \cite{bib:BM1,bib:BM2,bib:SJK,bib:iwamoto,bib:shukla3,bib:shukla4} and laboratory
\cite{bib:surko} variety), the fullerene plasma has a long enough life time that the collective
behavior  peculiar to the plasma state can be experimentally investigated under controlled
conditions.

The dynamics of pair plasmas is expected to be different from the
standard electron--ion plasma where the different mass of the
species, automatically breaks the symmetry between the
constituents. On the other hand, if the pair plasmas are prepared
in identical conditions, they must remain symmetric -- for example
their thermal speeds and temperatures are likely to be similar.

This is, indeed, the case as evident from the tremendous surge in
theoretical activity
\cite{bib:frank,bib:hans1,bib:H1,bib:H3,bib:shukla1,bib:shukla2}
to interpret and understand the experimental results
\cite{bib:OH1}. Most of these papers attempt to explain the
experimental findings in terms of the linear and nonlinear
properties that may be accessible to only pure pi plasmas.
Somewhat different problem appears to be when in PI plasma while
its creation a significant fraction of free electrons remain and
specific linear modes may develop in such plasmas
\cite{bib:H1,bib:H3}.

The aim of this paper, however, is totally different. We wish to propose, here, experiments that
explore phenomenon that are associated with slightly contaminated plasmas -- where the symmetry is
broken, say by the presence of a small amount of a heavier ions (heavier than the ions that
constitute the main plasma). Our motivation is two fold: 1) to study, per say, a possible nonlinear
bunching of electromagnetic waves through this new and exciting state of matter whose composition
may be highly controllable (naturally at some future date),  and  2) to create (from this
restricted perspective) an approximate replica of the cosmic plasma of MEV era which consists
primarily of electrons and positrons with a small concentration of symmetry-breaking ions
\cite{bib:BM1,bib:BM2,bib:shukla4}. The hope is that the laboratory experiment will show that,  in
a minimally contaminated  pair plasma, electromagnetism  can provide   mechanism for density
bunching lending further credence to the idea  of the electromagnetic origin of the large scale
structure of the universe.

It is important to stress that, although  cosmic considerations do
provide a motivation, the principal intent of this paper is to
investigate nonlinear phenomena in this potentially very
interesting, versatile, and controllable new state of matter.
Because of relatively low densities and heavier mass of the
species, the corresponding frequencies and scale lengths will be
in a novel range quite different from the e-p plasma counterpart.
One of the challenges, therefore, will be to derive laboratory
conditions suitable for the observance of bunching of the EM
waves. Similar behavior could be expected in  doped (or
dust-contaminated) fullerene plasmas in laboratory"
\cite{bib:shukla4}.

The fullerene plasmas of massive ions, however, are not quite
suitable for this experiment since all the frequencies associated
with the collective modes (plasma frequency, acoustic and Alfv\'en
frequencies) tend to be rather low.  Fortunately, the group of
Hatekayama and Oohara have already made considerable progress in
the production of the hydrogen, $H^+$--$H^-$  plasmas
\cite{bib:OH3,bib:OH4}. Since  the initial report,  both the
quality and quantity of this light pair ion plasma has  been
steadily improving. We will not dwell on how the $H^+$--$H^-$
plasmas are or will be produced and diagnosed (see
\cite{bib:OH3,bib:OH4}). We will simply assume that, in a not too
distant future, it will become  possible to "dial" in an
$H^+$--$H^-$ plasma  with densities sufficiently "high"  that
collective modes can be experimentally excited and studied
\cite{bib:Ha1}.

The main thrust of this paper is to investigate the nonlinear
interactions of EM waves with a primarily $H^+$--$H^-$ plasma
containing a small impurity of  high mass ions (positive or
negative). We will show that the symmetry breaking induced by the
immobile heavy ions (the main components, $H^+$ and $H^-$, will
have slightly different ambient densities  to insure charge
neutrality) leads to a finite electrostatic field  which, in turn,
makes  the plasma capable of supporting stable localized EM wave
structures of finite amplitude.

The pair ion plasma with a heavy ion contamination  will be
non-relativistic both in velocity and temperature as opposed to
the cosmic plasma it is supposed to mimic \cite{bib:BM1,bib:BM2}.
Although our eventual interest will be in the ($H^+$--$H^-$ +
heavy ion) plasma, we will analyze an arbitrary pair plasma whose
principal constituents have a mass $m$, and charge $\pm q$. The
equations of motion  for the dynamic species are:
\begin{equation}
\frac{d^{\pm}{\bf p}^{\pm}}{dt}+  \nabla {\hat{P}}^{\pm} =\pm
q\left[ {\bf E}+ \frac{1}{c}({\bf u}^{\pm}\times {\bf B})\right] ,
\label{eq:F-1}
\end{equation}
where \ $q=Z\,|e|$ ; \ ${\bf p}^{\pm}=m{\bf u}^{\pm}$ (${\bf
u}^{\pm}$) are the hydrodynamic momenta (velocities) of  the
oppositely charged species, \ ${\bf E}$ and ${\bf B}$ are the
electric and magnetic fields and $d^{\pm}/dt={\partial}/{\partial
t}+{\bf u}^{\pm}\cdot \nabla$ \ are the co-moving derivatives. In
Eq. (\ref{eq:F-1}), we have invoked a generic equation of state
$P^{\pm}=P^{\pm}(n^{\pm})$ to write the pressure term $(1/n^{\pm})
\nabla {P^{\pm}}$ as $\nabla {\hat{P^{\pm}}}$.

It has been shown earlier \cite{bib:BM1,bib:BM2} that: (a) the
perfectly symmetric system (\ref{eq:F-1}) cannot sustain a
localized electrostatic field, and (b) the electrostatic field is
essential for the localization of an electromagnetic wave passing
through a pair plasma. Since the creation of localized nonlinear
electromagnetic structures is the theme of this paper, a mechanism
for symmetry breaking must be provided. We do this by doping the
pi plasma with a heavy ion impurity (for the electrostatic mode
excitations in e-p-i plasmas see \cite{bib:shukla4}) so that the
demands of charge neutrality
\begin{equation} q\,n^+=q\,n^-+q^{\,'}N_{\infty}
\label{eq:F-2}
\end{equation}
will cause a difference between the densities $n^\pm$ of the main
constituents. Here $q'$ is the charge of the heavy, non--dynamic,
uniformly distributed, contaminating specie, and $\infty $ is
taken to be the fiducial point where the perturbations vanish.
Equation (\ref{eq:F-2}) implies the relation \
$q\,n^+_{\infty}=q\,n^-_{\infty}+q^{\,'}N_{\infty}$ for ambient
densities; it also leads to
 \begin{equation}
n^+-n^-=n^+_{\infty}-n^-_{\infty} \ , \label{eq:F-3}
\end{equation}

By design, the density $N_{\infty}$ of the contaminant is much
smaller than the density of the main pair  $n^{\pm}_{\infty}$
($N_{\infty}\ll n^{\pm}_{\infty}$), which, evolve via the
continuity equation:
\begin{equation}
{\partial n^{\pm}\over \partial t}+\nabla \cdot (n^{\pm} {\bf
u}^{\pm})=0.  \label{eq:F-4}
\end{equation}

The equations of motion, now, must be coupled with the Maxwell
equations. In terms of the vector (${\bf A}$) and the
electrostatic ($\phi$) potentials, the latter may be written as
(Coulomb gauge $\nabla\cdot {\bf A}=0$):
\begin{equation}
{\partial^2 {\bf A}\over \partial t^2}-c^2 \triangle {\bf
A}+c\,{\partial\over
\partial t}(\nabla \varphi)=4\pi\,c\,{\bf J},  \label{eq:F-5}
\end{equation}
\begin{equation}
\triangle \varphi=-4\pi\rho, \label{eq:F-6}
\end{equation}
where  the charge and current densities are defined by:
\begin{equation}
\rho=q(n^+-n^-)-q^{'}N_{\infty} ; \qquad {\bf J}=q(n^+{\bf
u}^+-n^-{\bf u}^-) . \label{eq:F-7}
\end{equation}

In terms of the  dimensionless variables
\begin{equation}
{\bf u}^{\pm}=\frac{{\bf u}^{\pm}}{c} , \
n^{\pm}=\frac{n^{\pm}}{n_{\infty}^{\pm}} , \ \hat{A}=\frac{q
\hat{A}}{mc^2} , \ {\bf r}=\frac{\omega_-}{c}{\bf r} , \
t=\omega_-t  , \label{eq:F-8}
\end{equation}
where $\hat{A}\equiv [{\bf A}; \varphi ; (q^{-1}T^{\pm})]$ and
$\omega_-=(4\pi q^2\,n_{\infty}^-/m)^{1/2}$ is the Langmuir
frequency of major negative species, the  defining equations read
(the continuity equations retain their form (\ref{eq:F-4})):
$$
{\partial^2 {\bf A}\over \partial t^2}-\triangle {\bf
A}+{\partial\over
\partial t}(\nabla \varphi)+
$$
\begin{equation}
+[n^-{\bf u}^- -(1+\epsilon)n^+{\bf u}^+]=0 \ , \label{eq:F-9}
\end{equation}
\begin{equation}
\triangle \phi=[n^--(1+\epsilon)\,n^+ +\epsilon \, ] \ ,
\label{eq:F-10}
\end{equation}
\begin{equation}
\frac{\partial {\bf \Pi}^{\pm}}{\partial t}={\bf u}^{\pm}\times
{\bf \Omega}^{\pm}-\nabla \psi^{\pm} , \label{eq:F-11}
\end{equation}
where the equations of motion (\ref{eq:F-1}), after standard
manipulation, have been rewritten in a revealing form that
contains  the generalized flows
\begin{equation}
{\bf \Pi}^{\pm}={\bf u}^{\pm}\pm {\bf A} ; \label{eq:F-12}
\end{equation}
the generalized vorticities
\begin{equation}
{\bf \Omega}^{\pm}=\nabla \times {\bf \Pi}^{\pm};\label{eq:F-13}
\end{equation}
and the effective energies
\begin{equation}
\psi^{\pm}={\hat{P}}^{\pm}+\frac{({\bf u}^{\pm})^2}{2}\pm \varphi
. \label{eq:F-14}
\end{equation}
The symmetry-breaking small parameter $\epsilon=
|q^{'}|N_{\infty}/qn_{\infty}^-\ll 1$, appearing in equations
(\ref{eq:F-9})--(\ref{eq:F-10}), will eventually be the source as
well as the measure of the electrostatic field.

We are, now, all set to analyze the one-dimensional propagation
(${\partial/
\partial z}\neq 0, {\partial/ \partial x}=0, {\partial/
\partial y}=0$) of a circularly polarized EM wave with a mean
frequency $\omega_o$ and a mean wave number $k_o$,
\begin{equation}
{\bf A}_{\perp}={1\over 2}({\hat{\bf x}}+i{\hat {\bf y}})\,A(z,t)
\exp\,(ik_o z-i\omega_o t) +c.c.\label{eq:F-15}
\end{equation}
Here ${\hat{\bf x}}$ and ${\hat{\bf y}}$  are the standard unit
vectors, and the envelope $A(z,t)$ is a slowly varying function of
$z$ and $t$. Using the gauge condition ({$A_z=0$ in the present
context}), we can calculate
$$
{\bf u}^{\pm}\times {\bf \Omega}^{\pm}={\bf u}^{\pm}\times ({\bf
\hat{z}}\times \frac{{\partial \bf \Pi}^{\pm}_{\perp}}{\partial
z})=
$$
\begin{equation}
=\hat{\bf z}\,{\bf u}^{\pm}\cdot \frac{{\partial \bf
\Pi}^{\pm}_{\perp}}{\partial z}-u_z^{\pm}\frac{{\partial \bf
\Pi}^{\pm}_{\perp}}{\partial z} \ , \label{eq:F-16}
\end{equation}
and substitute it in the transverse component of
(\ref{eq:F-11}) to derive:
\begin{equation}
\left(\frac{\partial}{\partial t}+u_z\frac{\partial}{\partial z
}\right){\bf \Pi}^{\pm}_{\perp}=0 \ .   \label{eq:F-17}
\end{equation}
The simplest solution ${\bf \Pi}^{\pm}_{\perp}=0$ is the most
relevant here since, for the localized solutions we are seeking,
all fields must go to zero at infinity. The consequential relation
\begin{equation}
{\bf u}^{\pm}_{\perp}=\mp {\bf A}_{\perp} . \label{eq:F-18}
\end{equation}
relates the transverse components of the hydrodynamic velocities
and the vector potential.

The longitudinal dynamics is obtained  from the $z$ component of
(\ref{eq:F-11}) using (\ref{eq:F-16}). Noting that all terms in
this dynamics vary on a slow time scale, we can introduce the
following variables for convenience: $\xi =z-v_gt \ , \tau =t$ ,\
where $v_g=k_0/\omega_0$ \ is the group velocity of the EM wave
packet. Assuming $v_g\partial/\partial \xi\gg
\partial/\partial \tau$, and integrating, we find:
\begin{equation}
u^{\pm}_z=\frac{1}{2}\,{u^{\pm}_z}^2+\frac{1}{2}\,A_{\perp}^2\pm
\phi + (\hat{P}^{\pm}-\hat{P}^{\pm}_{\infty}) \ , \label{eq:F-19}
\end{equation}
The result holds in  the transparent plasma limit ($\omega_0\gg
1$) for which $v_g\simeq 1$. In the laboratory experiments, this
condition, which in dimensional terms demands {$\omega_0\gg
\sqrt{2}\,\omega_-$ ($=\omega_p$ the plasma frequency)} can be
easily arranged for pi plasmas \cite{bib:OH1,bib:OH3,bib:OH4}. In
the non-relativistic case under consideration (the velocities are
normalized to $c$), the first term on the r.h.s. of
(\ref{eq:F-19}) is readily neglected. However the second term
$A_{\perp}^2$= ${u^{\pm}_{\perp}}^2$ must be kept because
${u^{\pm}_{\perp}}^2$ is allowed to be $\gg {u^{\pm}_z}^2$. In
(\ref{eq:F-19}), the constants of integration are determined from
the boundary condition  that, at infinity,  EM fields and the
plasma momenta vanish.

Similarly, the continuity equation yields:
\begin{equation}
n^{\pm}=\frac{1}{1-u^{\pm}_z}\simeq \left(1+u^{\pm}_z\right) .
\label{eq:F-20}
\end{equation}
From (\ref{eq:F-19}) and (\ref{eq:F-20}), and the quasi-neutrality
condition (obtained  by neglecting $\triangle \phi$ in the Poisson
equation (\ref{eq:F-10}) one derives, after straightforward
algebra, the following major relationships:
\begin{equation}
\phi \simeq \frac{\epsilon}{4}\,|A_{\perp}|^{\,2} \
\label{eq:F-21}
\end{equation}
and
\begin{equation}
u_z^-\simeq 2\epsilon^{-1}\,\phi , \quad n^++n^-\simeq
2\,(1-\epsilon \ \phi ) . \label{eq:F-22}
\end{equation}
As expected, the electrostatic field (that will bring in the
nonlinearity, necessary for localization, in the Maxwell equation
(\ref{eq:F-9})) owes its existence to  symmetry breaking, and is
proportional to the parameter $\epsilon \ll 1$. It is clear that
the electrostatic potential will be absent in pure pair plasmas
with no contamination. This phenomenon was, first, investigated
for a relativistic plasmas in a cosmic setting \cite{bib:BM1}. In
a pure pair plasma (equal density and  equal temperature for the
two species) the radiation pressure imparts equal longitudinal
momenta to both the negative and positive ions (since their
effective masses are equal) and, thus, causes no charge separation
($n_-=n_+$ and $\phi=0$).

To proceed further,  let us determine the transverse current
defined in (\ref{eq:F-7}) or (\ref{eq:F-9}) by resorting to
(\ref{eq:F-18}), (\ref{eq:F-21}), and(\ref{eq:F-22}). To the
lowest order in the nonlinearity, we find,
\begin{equation}
{\bf J}_{\perp} =-\left[(2+\epsilon)-\frac{1}{4}\ \epsilon^2
\,|A_{\perp}|^2\right]\,{\bf A}_{\perp} \ . \label{eq:F-23}
\end{equation}
Notice that the nonlinear term coincides with the non relativistic
limit of the expression derived in \cite{bib:BM2}. We see, that
without an asymmetry ($\epsilon \equiv 0$), the nonlinear term
disappears since the electrostatic potential (responsible for this
term) $\phi $ is now zero.

When the transverse current, evaluated in (\ref{eq:F-23}), is
substituted into (\ref{eq:F-9}), we obtain the  nonlinear equation
for the evolution of the  slowly varying envelope function $A$.
Following the standard methodology (see \cite{bib:BM2}, for
example), we can derive the final equation in the parabolic
approximation:
\begin{equation}
2i\omega_0\frac{\partial A}{\partial \tau
}+\frac{2}{\omega_0^2}\frac{\partial^2 A}{\partial z^2}
+\frac{1}{4}\,\epsilon^2\,|A|{\,^2} \,A=0 \ , \label{eq:F-24}
\end{equation}
where the wave frequency $\omega_0$ satisfies the dispersion
relation \ $\omega_0^2=(2+\epsilon )+k_0^2$ \ implying $v_g\simeq
1$ for a transparent plasma for which $\omega_0\gg 1$ has been
assumed. With the additional obvious normalization, Eq.
(\ref{eq:F-24}) is cast into the standard form of the Nonlinear
Schr\"odinger equation (NLSE),
\begin{equation}
i\,\frac{\partial A}{\partial \tau }+\frac{\partial^2 A}{\partial
z^2}+ \epsilon^2|A|^2\, A=0 . \label{eq:F-25}
\end{equation}

It is well known that the NLSE  can be solved for solitons by the
inverse scattering method. The usual stationary soliton solution
is obtained by letting ($\Omega$ is a constant that
corresponds to a nonlinear frequency shift)
\begin{equation}
A=A(\xi)\,e^{i\Omega^2\tau}    \label{eq:F-26}
\end{equation}
that leads to the nonlinear ordinary differential equation(ODE)
\begin{equation}
\frac{\partial^2 A}{\partial
\xi^2}-\Omega^2\,\left[{1-\frac{\epsilon^2}{\Omega^2}\,A^2}\right]\,A=0.
\label{eq:F-27}
\end{equation}
With boundary conditions appropriate to a localized
solution, $A=0=dA/d\xi$ as $\xi \to \pm\infty $, Eq. (\ref{eq:F-27}) yields
\begin{equation}
A=A_m\,{\rm sech}{\left({\frac{\epsilon
A_m}{\sqrt{2}}\,\xi}\right)} , \label{eq:F-28}
\end{equation}
a lump of the  electromagnetic field potential with amplitude
$A_m=\sqrt{2}/\epsilon$, and localization width
$d$=$\sim\sqrt{2}/\epsilon A_m=1$.  In standard measures,
therefore, the localization width is the collision less  skin
depth of the main plasma ($d\sim\lambda_{skin}$).

We have, thus demonstrated that the pair ion plasmas for which the
symmetry is broken by a slight contamination (doping) of a heavier
immobile ion can support stable localized EM wave structures even
in the non relativistic limit appropriate to the current and near
future laboratory experiments. To appreciate the importance of
theoretical results in the context of a laboratory setting, let us
invoke a hypothetical $H^+$--$H^-$ plasma of density
$10^{10}\,cm^{-3}$ that is expected to become available in near
future \cite{bib:OH4,bib:Ha1}. For these densities, the
characteristic plasma frequency $\omega_p\simeq
1.86*10^{8}\,s^{-1}$ corresponding to a
$\nu_p=\omega_p/{2\pi}\simeq 30\,MHz$. The first condition for our
derivation that the plasma be transparent ($\omega_0\gg \omega_p$)
can be easily satisfied by choosing the EM wave frequency to be in
the few hundred MHz range.

Since the  localization distance was shown to be about a skin
depth, for $\omega_p= 1.86*10^{8}\,s^{-1}$, it comes out to be
$d\sim\lambda_{skin}=c/\omega_p\sim 1.6\,m$. Thus to
experimentally observe density and  electromagnetic field bunching
in a doped $H^+$--$H^-$ plasma, the plasma must  extend over
several meters (along the propagation direction). Naturally, for
higher densities, the localization distance will decrease and the
length requirement on the experiment will be  correspondingly
relaxed.

\bigskip

By investigating the nonlinear propagation of high intensity EM
waves in a pair ion plasma contaminated with a small fraction of a
high mass immobile ions (for symmetry breaking), we have
highlighted a very remarkable property of this new and interesting
state of (laboratory created) matter -- it can strongly localize
the EM radiation with finite density excess in the region of
localization. It must be stressed that this is just one example of
a vast variety of physical phenomena (many with astrophysical
consequences) associated with pair plasmas that can be explored in
controlled experiments.

It is worth repeating that the equation (\ref{eq:F-25}) obeyed by
the field envelope $A$ comes out to be the NLSE which is known to
have exact soliton solutions. Solitons carry a large density
inhomogeneity, and form non diffractive, non dispersive localized
structures. Such objects  have been called "heavy bullets of
light", and  their possible important role in cosmology as a
source of structure formation has been already alluded to
\cite{bib:BM1,bib:BM2}. The solitons are also important for energy
transfer investigations in plasmas.

There are several ways in which this simple idealized calculation
can be extended and augmented. For example we could, $\it inter$ $
\it alia$: 1) investigate other modes of symmetry breaking, 2)
introduce a guide magnetic field (see \cite{bib:shukla4} for
electrostatic mode envelope excitations in pair-ion plasmas doped
by massive ions (or dust particles), and 3) increase the
dimensionality by introducing weak transverse dependence.

\bigskip

The authors acknowledge special debt to the Abdus Salam
International Centre for Theoretical Physics, Trieste, Italy. The
work of SMM was supported by USDOE Contract No.DE-FG03-96ER-54366.
The work of NLS was partially supported by ISTC Project G-1366 and
Georgian NSF grant projects GNSF 69/07 (GNSF/ST06/4-057) and GNSF
195/07 (GNSF/ST07/4-191).

\end{document}